\documentclass[amsmath,amssymb, floatfix]{revtex4}

\usepackage{graphicx}
\usepackage{dcolumn}
\usepackage{bm}
\usepackage{amssymb}
\usepackage{amsmath}

\begin{document}

\title{Prediction in the Hypothesis-Rich Regime}

\author{Andr\'{e} X. C. N. Valente$^{1,2}$}
\address{
$^{1}$Centro de Neuroci\^{e}ncias e Biologia Celular, Universidade de Coimbra, Portugal\\
$^{2}$ Unidade de Sistemas Biol\'{o}gicos, Biocant, Cantanhede, Portugal\\
}


\begin{abstract}
We describe the fundamental difference between the nature of problems in traditional physics and that of many problems arising today in systems biology and other complex settings.
The difference hinges on the much larger number of a priori plausible alternative laws for explaining the phenomena at hand in the latter case.
An approach and a mathematical framework for prediction in this hypothesis-rich regime are introduced. 
\end{abstract}


\maketitle

\section{Introduction}
Science can be seen as an attempt to find good laws, rules that given some observed input information, are able to predict accurately a resulting, yet unobserved, outcome of interest~\cite{OtherReq}.
Experimental observations guide this search, by allowing proposed laws to be matched against a body of data where both input information and respective outcome are available.
Even in traditional physics problems, it is important to realize that generally there exist a multiplicity of laws that replicate well the available experimental data.
For example, planetary motion is very accurately predicted by Newton's inverse square gravitational law.
But consider an alternative law stating that this is true up to the year 2100 AD, beyond which the force decreases cubically, rather than quadratically with the distance.
Based solely on matching available planetary motion experimental data, this other law is exactly as good as Newton's law.
Similarly, as shown by both Ptolemy and Copernicus, it is possible to construct an epicycle based theory that correctly predicts available planetary motion data~\cite{Epicycle}.
The reason we dismiss these alternatives is that, unlike Newton's law, we consider them irrational.

{\em The objective is therefore to find a law that i) is rational and ii) matches available data.}
Contributors to determining the rationality of a law include both a) information beyond the data in item ii) above, from observations not directly related to the predictions of the law and b) fundamental beliefs that we hold to be true.
For instance, rationality requires assuming properties such as the underlying homogeneiety of space and time.
Minimal good-behavior of any involved mathematical functions is generally expected in a rational law.
Laws based on conservation of some mathematically defined quantity are often considered particularly rational too.
As a final example, the simplicity of a law (relative to the problem at hand) is usually equated with its rationality, i.e., the Occam's Razor principle.

The nice property of most typical problems in physics is that, once an arguably rational law that matches the observed data is found, we can be fairly confident that it will predict equally well the yet unobserved Cases.
This is because in these problems the rationality constraint we are imposing is restrictive enough that it is very unlikely that an arguably rational law would match available experimental data just by chance.
However, in contrast with the above situation, the typical systems biology~\cite{SysBio} problem involves many a priori plausibly relevant variables and generates many more rational alternative hypothesis.

Let us introduce one possible formalization of these issues (Fig. 1).
Define a finite measure~\cite{Rudin} on the Space of Cases of Interest and a metric on the Space of Outcomes.
Together, these induce a metric on the Space of Laws (a space of mappings from the Space of Cases of Interest to the Space of Outcomes). 
Call it the True Metric~\cite{TrueMetric}. 
Let the idealized Perfect Law map every Case to exactly its correct outcome.
Now associate with each law a {\em True Distance}, the distance between that law and the Perfect Law under the True Metric.
The True Distance of a law gives an averaged distance between its predicted outcomes and the correct outcomes. 
The True Distance of a law is unknown to us.
Our ultimate objective is to find a law with a low True Distance (with the practical restriction of mapping all Cases with identical observed input information to the same Outcome).
Define an {\em Observed Distance}, analogously to the True Distance, except in that: i) It is based solely on the Cases with available experimental data~\cite{ObservedMetric} and ii) the term of comparison is not the correct outcomes of the Perfect Law, but  rather those experimentally measured outcomes.
The Observed Distance of a law gives an averaged distance between predicted and observed outcomes for the set of Cases with experimental data.
Observed Distance is defined to provide the best possible {\em computable} estimate of True Distance, based {\em exclusively} on the available experimental data (i.e., rationality considerations aside).
Each specific scientific problem may require its more particular, tailor-made, Observed Distance definition~\cite{Mezic}.
For our purposes, only assessing the relative True Distances of laws will prove relevant.

The body of available experimental data is the realization available to us of its corresponding random variable.
The computed Observed Distance of a law is therefore also the available realization of its corresponding random variable.
Since usually the experimentally observed data covers but a small subset of the Space of Cases of Interest and, additionally, the observed outcome measurements can be noisy, Observed Distances are going to be off from the respective True Distances.
Therefore, if the set of laws we consider rational is too large, in particular if it contains too many laws with a high True Distance, then, statistically, likely there will be high True Distance laws with Observed Distances as low or lower than the Observed Distances of the low True Distance laws we would like to select out (Fig. 2).
This difficulty is the so called over-fitting, or under-determination of a problem~\cite{Overfit2}.
Unlike in traditional physics, the nature of problems arising in systems biology and other complex settings~\cite{SystemsMed,Urban, CmplxNets} is such that at least the possibility of this occurring tends to be present.
 We next introduce an approach for doing science in this fundamentally different hypothesis-rich regime.
 
\section{Rationality classes}
Let the final aim be to select a single law, with a True Distance as low as possible.
Computational time considerations in the search and testing of candidate laws are disregarded. 
We i) define a Set of Candidate Laws based on rationality considerations (defined exclusive of the information in the experimental data used in the Observed Distance computation) and then ii) select the law from this set with the lowest Observed Distance.
The likely quality (True Distance) of the selected law is determined by the {\em True Distance Distribution and Correlation Structure} (henceforth shorthanded TDD) of the Set of Candidate Laws~\cite{Structure}.
Namely, it is determined by what True Distance values are statistically likely to be prevalent at the lowest Observed Distance (Fig. 3, Basic Method).
The search for a good law therefore can be equated with the search for a set of laws with a favorable TDD.
The set of all possible laws will usually have a terrible TDD that leads to the selection of a bad law.
Defining a rational Set of Candidate Laws is an attempt to obtain a more favorable TDD.

Typically a variety of presumed equally rational arguments can be put forward to generate a rational set of laws.
Call the set of laws generated by each such argument a Rationality Class.
As an example, consider prediction of tumor growth/evolution~\cite{CancerBook, CancerRealistic, CancerReaction}.
Suppose a particular set of biological considerations about cancer growth leads to a model with seven partial differential equations with five undetermined parameters.
This model defines a Rationality Class, each set of values for its parameters giving rise to a law in that class.
Experimental data taken on the evolution of a number of actual tumors could then be used to chose the best values for these five parameters.
This is the process of selecting the law with the lowest Observed Distance in the class.
But perhaps additional biological arguments could be raised, that if true would either settle some of these parameters or constrain their possible range. 
The model under this restricted parameter space would define a second Rationality Class.
On the other hand, arguably, cancer growth prediction is really best if modeled at a more detailed scale.
This could lead to a set of thirty partial differential equations with twenty undetermined parameters.
This would constitute yet another distinct Rationality Class.

Most importantly, even though we are unable to determine a priori which of the Rationality Classes is the best, we know their distinct origin gives them a fair chance of having significantly different TDDs.
A key insight is that, in such cases, merging these classes into a single Set of Candidate Laws and selecting the law with the lowest Observed Distance is often not ideal, as that law may actually not be coming from the class with the most favorable TDD.

We therefore propose a two-level hierarchical process for selecting the single final law (Fig.~3, Method of Rationality Classes).
This process entails the division of the available experimental data into two separate test sets.
First, from each Rationality Class, the law with the lowest Observed Distance based on the first test set is selected.
The selected laws are then compared in the second step, the one amongst them with the lowest Observed Distance based on the second test set being picked as the final choice.
Such a two-level process, permits effective comparison of the solutions produced by different Rationality Classes.
The cost of the process is that the experimental data has to be divided into two separate smaller sets, therefore lowering in each case the accuracy of Observed Distances as estimates of True Distances.
This cost can be worthwhile as some of the Rationality Classes may have substantially different TDDs.
If the classes had been merged into a single Set of Candidate Laws, lower Observed Distance laws from a Rationality Class with an unfavourable TDD could occlude a higher Observed Distance  but better (lower True Distance) law from a Rationality Class with a more favourable TDD.
Note how the goal has moved to one of identifying a collection of Rationality Classes worth testing (hoping that the collection will contain a good class, as it in turn will yield a good final law, picked in the second step of the above process).

Consider a Rationality Class that cannot be broken down via some argument into distinct Rationality Classes. 
Laws in this class that further have the same Observed Distance are completely indistinguishable.
In such a class, by definition, it is impossible to do better than to select the law with the lowest Observed Distance.
In particular, no random breaking of the class into separate classes can overcome this fundamental limitation.
Also, it is always best to test all the laws in a given Rationality Class.
Not doing so is tantamount to not selecting the lowest Observed Distance law in that class.

In principle, it is possible to keep creating Rationality Classes, or to keep dividing a Rationality Class into more and more separate classes via gradually subtler arguments.
Recall that each of these classes sends its selected law to the second level of the process.
Together, these selected laws form their own class at the second level, with its own TDD, from where the final law is selected.
The discussion pertaining to the TDD of a class is equally valid for this new class at the second level.
Therefore, in particular, a non-judicious, too liberal class creation at the lower level could result in an unfavourable TDD at the second level.

\section{Recursive framework}
We now briefly introduce a recursive framework (Fig. 4) that optimizes some aspects of this new approach to prediction based on Rationality Classes.
First, note that it may be profitable to extend the concept of Rationality Classes to higher-level grouping of Rationality Classes and so fourth, as shown in the Hierarchical Tree Structure of Figure 4.
This is so as, analogously to what happens with basic Rationality Classes, the Higher-Level Rationality Classes (at some given higher-level) may have distinct TDDs and therefore considering them as a single Class may be detrimental. 
Second, the Figure 4 recursive framework helps extract more out of the set of Cases with available outcome data.
 It does so, i) via the repeated splitting of T into different (T*, U*) combinations when determining the best child node to select at any given generic node $x$ and ii) via finally passing to that selected child node (Best of the Child Nodes recursive call) the full set T originally received by the parent  generic node $x$.
Note how Cases in the input set U of unknown Cases to a node $x$ never become cases in the input  Set T of available Cases to a downstream node, called within a recursive thread originating at $x$.
Thus, as proper, the process to generate outcome predictions for the set U never makes use of the set U of outcomes.

\section{Complex Biomarkers}
Although most pathological conditions are multifactorial in nature, the search for the single entity Biomarker~\cite{BiomarkerDef} has been a dominant paradigm in medicine~\cite{Combinatorix}.
Methods for detecting protein interactions~\cite{Interactome1, Interactome2}, for assessing transcriptional activity~\cite{Microarray} and for synthetic lethality screening~\cite{Synthetic} are examples from a rapidly expanding collection of techniques for acquiring biological data in a high-throughput, quantitative manner.
An open question is whether integrated analysis of such data can lead to the identification of Complex Biomarkers that, based on multiple biological readings,  more accurately assess the system-level status of a biological process.
We view blunt integration of data into a single statistical model as a nonideal approach for finding a Complex Biomarker:
It amounts to testing only one Rationality Class (Fig. 3, Basic Method).
On the other hand, through appropriate biological hindsight, the multiple data sources may prove ideal to create a diversity of Rationality Classes for testing within a framework as outlined in this article.

In Valente et al.~\cite{PD}, we use our approach to analyze blood genome-wide expression data from 50 Parkinson's disease patients and  55 age-matched controls~\cite{Scherzer}. Specifically, a classifier law for distinguishing Parkinson's disease individuals from non-Parkinson's disease individuals based on blood expression data is constructed. An interesting result of the analysis is that it hints at differential expression in blood cells in many of the processes known or predicted to be disrupted in Parkinson's disease. Starting from this observation, a hypothesis on Parkinson's disease originating in a hematopoietic stem-cell differentiation process expression program defect is proposed.

\section{Summary}
In this article we put forward that problems in systems biology and other complex settings often fall in the hypothesis-rich regime.
We argue that doing science in this regime requires a fundamentally distinct approach, which is proposed.
Namely, the concept of Rationality Classes is introduced and the focus is shifted from one of looking for the right solution to one of identifying favorable Rationality Classes.
Finally, we present a recursive framework that attempts to maximize prediction power under this new approach.

\newpage
\bibliographystyle{elsarticle-num}

\begin{thebibliography}{99}
\bibitem{TrueMetric}
Define the True Metric distance between two laws $f$ and $g$ as $\int_\text{All space} d\big(f(\!x\!),g(x)\big)d\mu$, where $x$ is the Case variable, $\mu$ is the finite measure over the Space of Cases of Interest and $d\big(f(\!x\!),g(x)\big)$ is the metric distance between $f(x)$ an $g(x)$ in the Space of Outcomes.
\bibitem{ObservedMetric}
Now the measure used in the Space of Cases of Interest assigns only nonzero point masses to the Cases with experimentally observed outcomes. 
As defined, this measure is appropriate only in a setting where the set of Case observations is finite.
\end{thebibliography}

\newpage
\begin{figure}
\vspace{2cm}
\hspace{0cm}\scalebox{.48}{\includegraphics{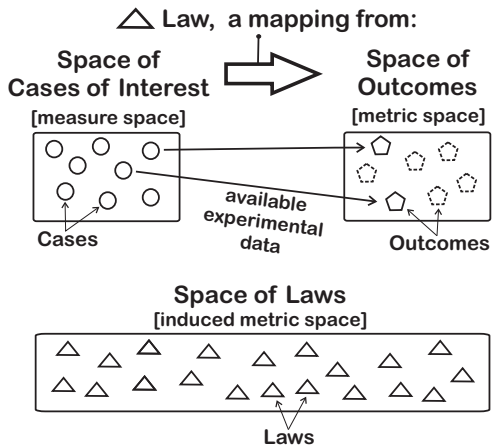}}
\caption{
A formalization of the scientific problem.
The Space of Cases of Interest is assumed fixed a priori.
Each Case has associated with it a correct outcome and some experimentally observed input information.
Measurement noise is handled by considering as distinct Cases even Cases that differ only in the observed input information.
The goal is to find a law that can accurately predict a Case outcome, given the Case input information (i.e., subject to mapping all Cases with the same observed input information to the same outcome).
To help, there is a body of (noisy) data on the outcomes of some Cases.
}
\end{figure}

\begin{figure}
\vspace{3cm}
\hspace{0cm}\scalebox{.55}{\includegraphics{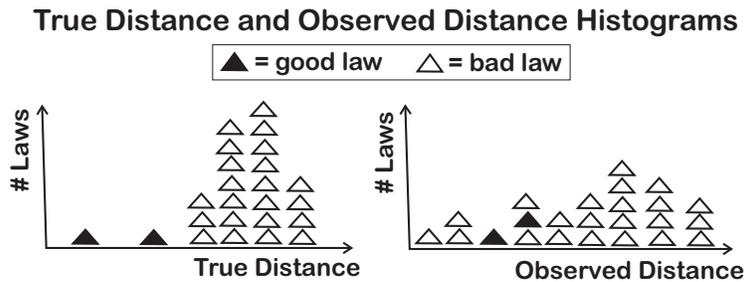}}
\caption{The presence of too many (bad) laws with a high True Distance (left chart) will, Observed Distance wise, occlude the low True Distance (good) laws (right chart).}
\end{figure}
\newpage

\begin{figure}
\vspace{3cm}
\hspace{0cm}\scalebox{.69}{\includegraphics{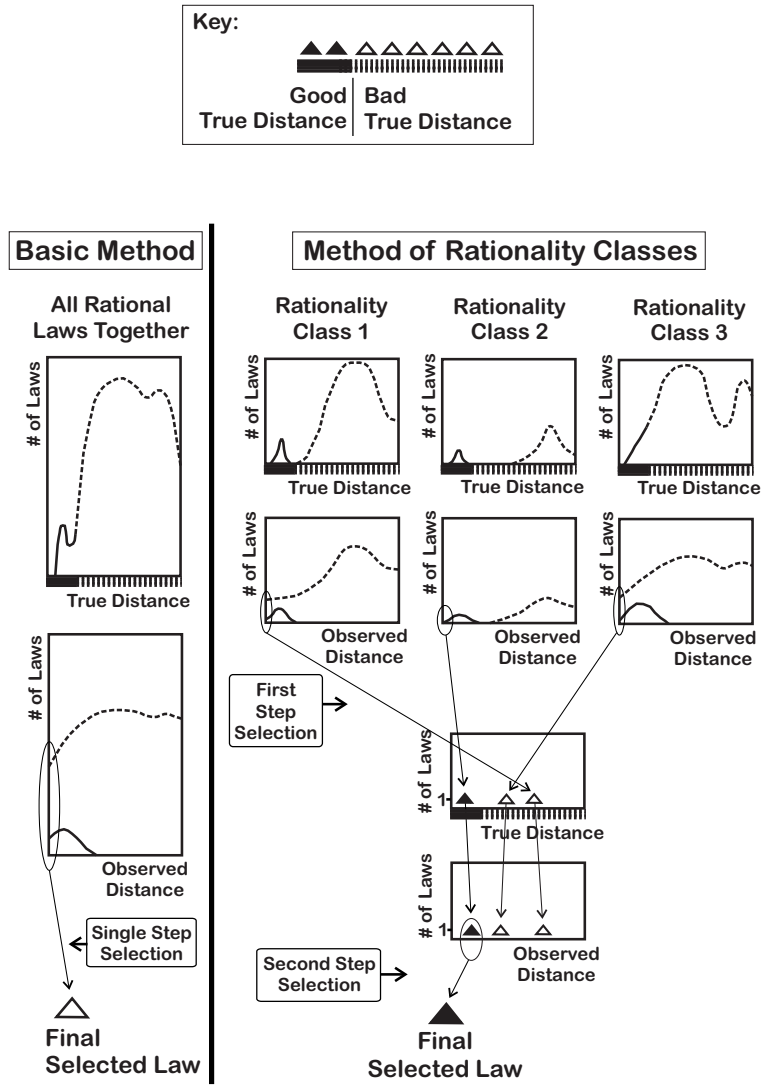}}
\caption{
Two alternative methods for selecting a final law.
For distinction, solid lines denote good True Distances, dashed lines denote bad True Distances.
Similarly, a solid triangle represents a law with a good True Distance, an open triangle represents a law with a bad True Distance. 
{\em Basic Method:} a rational Set of Candidate Laws is defined. 
Its True Distance Distribution and Correlation Structure is unknown to us (\# of Laws vs. True Distance histogram; note that the correlation structure is not shown).
The Observed Distances of these laws, generated by the available experimental data, are known.
The \# of Laws vs. Observed Distance histogram shows a typical Observed Distance distribution for the good True Distance laws and similarly for the bad True Distance laws.
The final law is selected from those with the lowest Observed Distance.
In this example, a bad True Distance final law was selected, as most of the laws at the lowest Observed Distance had a bad True Distance.
{\em Method of Rationality Classes:} 
The same rational Set of Candidate Laws this time is divided into 3 Rationality Classes, each with its own True Distance Distribution and Correlation Structure.
The available experimental data is divided into two test sets.
Based on the first test set, a lowest scoring Observed Distance law is selected from each of the three classes.
Based on the second test set, new Observed Distances are generated for these three laws, the one with the lowest one being selected as the final law.
In this example, the existence of a  favorable True Distance Distribution and Correlation Structure in Rationality Class 2, together with the two-level hierarchical procedure, allowed a good True Distance law to be sifted as the final selected law.
}
\end{figure}

\newpage

\begin{figure}
\vspace{3cm}
\hspace{0cm}\scalebox{.53}{\includegraphics{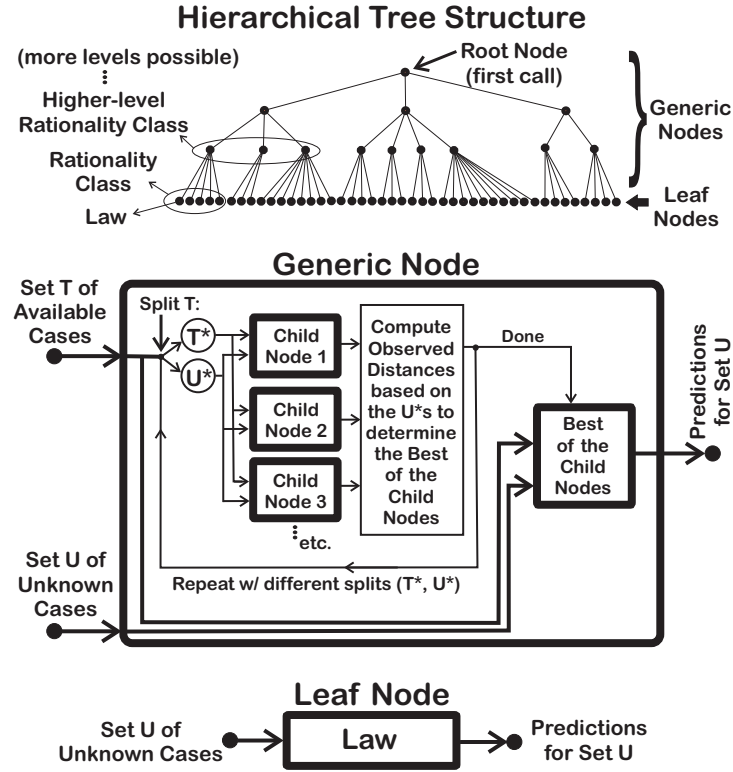}}
\caption{
Recursive framework for prediction in the hypothesis-rich regime.
{\em Hierarchical Tree Structure:} Laws are grouped into Rationality Classes, which in turn are grouped into Higher-Level Rationality Classes and so forth.
{\em Generic Node:} A generic node $x$ receives as input i) a set T of Cases with available outcome data for its use and ii) a set U of unknown Cases, where the outcome data is not provided to the node $x$ (even if known).
Node $x$ outputs predictions for the outcomes of Cases in this set U.
Inside any given generic node $x$, recursive calls can be made to child nodes of node $x$ (nodes directly linked to node $x$ in the Hierarchical Tree Structure diagram; orange boxes inside the Generic Node diagram).
Note that each node gives rise to multiple recursive threads.
The repeated splitting of T into (T*, U*) can be implemented using any standard cross-validation or resampling scheme~\cite{CrossValidation} (among other issues, the choice of scheme will have to factor in computational time costs, something not addressed in this article).
{\em Leaf Node:} a recursive thread ends when a leaf node is reached. In a leaf node, a law generates predictions for the Cases in U it receives.
Note that in the generic nodes that recursively call leaf nodes, T simply becomes U*, as laws have no use for the T* set.
The whole process starts at the root node and ends when the root node produces predictions for its set U of unknown Cases.
}
\end{figure}

\end{document}